\documentclass[conference]{IEEEtran}
\IEEEoverridecommandlockouts
\usepackage{cite}
\usepackage{amsmath,amssymb,amsfonts}
\usepackage{algorithmic}
\usepackage{graphicx}
\usepackage{textcomp}
\usepackage{xcolor}
\usepackage{hyperref}

\def\BibTeX{{\rm B\kern-.05em{\sc i\kern-.025em b}\kern-.08em
    T\kern-.1667em\lower.7ex\hbox{E}\kern-.125emX}}
\begin{document}

\newcommand{\alon}[1]{{\it\small\textcolor{purple}{[[[ {#1}\ --alon ]]]}}}
\newcommand{\vivian}[1]{{\it\small\textcolor{blue}{[[[ {#1}\ --vivian ]]]}}}

\title{Jo: The Smart Journal\\}

\author{\IEEEauthorblockN{Vivian Li\IEEEauthorrefmark{1}
, Alon Halevy\IEEEauthorrefmark{1}, Adi Zief-Balteriski Ph.D.\IEEEauthorrefmark{1}, Wang-Chiew Tan\IEEEauthorrefmark{1}, George Mihaila\IEEEauthorrefmark{1},
John Morales\IEEEauthorrefmark{1}, \\
Natalie Nuno\IEEEauthorrefmark{1}, 
Huining Liu\IEEEauthorrefmark{1}, Chen Chen\IEEEauthorrefmark{1}, 
Xiaojuan Ma\IEEEauthorrefmark{2}, Shani Robins Ph.D.\IEEEauthorrefmark{3},
Jessica Johnson\IEEEauthorrefmark{4}}
\IEEEauthorblockA{\IEEEauthorrefmark{1}Megagon Labs, Mountain View, USA\\
\{vivian,alon,adi,wangchiew,george,john,natalie,huining,chen\}@megagon.ai}
\IEEEauthorblockA{\IEEEauthorrefmark{2}Hong Kong University of Science and Technology, Hong Kong, China\\
mxj@cse.ust.hk}
\IEEEauthorblockA{\IEEEauthorrefmark{3}Stanford University, Stanford, USA\\
shani@wisdomtherapy.com}
\IEEEauthorblockA{\IEEEauthorrefmark{4}Sofia University, Palo Alto, USA\\
jessicaglobalchi@gmail.com}}

\maketitle

\begin{abstract}
We introduce Jo, a mobile application that attempts to improve user's well-being. Jo is a journaling application--users log their important moments via short texts and optionally an attached photo.  Unlike a static journal,  Jo analyzes these moments and  helps users take action towards increased well-being. For example, 
 Jo annotates  each moment with a set of values (e.g., family, socialization, mindfulness), thereby giving the user insights about the balance in their lives.  In addition,  Jo helps the user create reminders that enable them to create additional happy moments. We describe the results of fielding Jo in a study of 39 participants. The results illustrate the promise of a journaling application that provides personalized feedback, and points at further research. 

\end{abstract}


\section{Introduction}
Positive psychology is a field 
that studies the factors that sustain people's well-being over
time~\cite{seligman2011flourish,fredrickson2009positivity,lyubomirsky2008happiness}. 
One of its interesting findings 
is that while 50\% of our happiness is genetically determined, and only 10\% of it is determined by our
life circumstances (e.g., finances, job, material belongings),
40\% of our happiness is determined by behaviors that are
under our control~\cite{Diener:1999}. Examples of such behaviors include investing in long-term personal relationships, bonding with loved ones, doing meaningful work, and caring for one's body and mind. Consequently, positive psychologists have focused on devising methods to steer people towards those behaviors. 

Naturally, there has been significant interest to develop technologies that help people incorporate the findings of positive psychology into their daily lives. Currently, there are two categories of applications available.  The first  includes journaling applications that enable the user to write down their daily experiences,  accompanied by other media, such as photos and videos~\cite{Bliss,Mojo}. However, these applications offer little feedback to the user except for pointers to relevant reading material. The second category includes applications that give advice to users based on survey questions that are asked on a regular basis~\cite{Happify,Happier,Killingsworth}. However, the survey questions can be burdensome  and they are somewhat generic and therefore do not relate to the user's own nuanced experiences. 

We describe Jo, a smartphone application that combines the benefits of both approaches and provides feedback to the user based on their own experiences. Jo is a journaling application--users log their important moments via short texts and optionally an attached photo (e.g., ``Had great dinner with my parents", ``Enjoyed 5 mile run around the lake". In addition to providing the benefits of a journal, such as revisiting and reflecting on their experiences,  Jo analyzes these moments and  helps users take action towards increased well-being. First, Jo uses Natural Language Processing techniques to annotate each moment with a set of values (e.g., family, socialization, mindfulness, compassion for others), thereby giving the user insights about the balance in their lives.  Second, Jo helps the user create reminders that enable them to create additional happy moments. For example, if the user reports a good social interaction with a friend, Jo will ask if they want to create a reminder to meet with that friend again. Over time, Jo will learn to give more insights and suggestions on how the user can spend their time to increase their well-being. 

In the long run, creating a repository of personal moments can lead to significant benefits. 
The key challenge in  building Jo is to provide users with immediate feedback that they find useful in order to engage them with the system in the short term. To that end, this paper addresses  two technical challenges.  
The first challenge is to select the appropriate set of features and UI interactions that  provide users with immediate value as they start building their journal.  The second challenge is to develop the NLP techniques that enable Jo to provide specific feedback to the user based on their journal entries.

We make the following contributions in this paper. 
First,  we describe the design of Jo and its rationale. In particular, we describe  the values that Jo recognizes, how it helps in reflecting, and the feedback that Jo provides (Section~\ref{sec:jo-design}).   Second,  we describe the NLP  algorithms underlying Jo, including the  algorithm for classifying the affective polarity of moments, mapping moments to Jo's values, and how Jo recognizes activities in moments in order to give advice to the user (Section~\ref{sec:nlp}).  Finally, we describe the field studies we conducted with Jo. The main takeaways from the study is that users like a  journaling application that provides feedback and that with careful design of prompts to the user, we can get them to reflect on different aspects of their lives (Section~\ref{sec:evaluation}).

\section{Related work}
Several research prototypes and commercial applications have addressed issues related to Jo.
The Reflection Companion Project~\cite{Kocielnik:2018:RCC:3236498.3214273} is a mobile conversational system that provides daily adaptive dialogues to assist users' reflection. The feedback is based on their physical activity data collected by fitness trackers. The personalized feedback was effective in sustaining user engagement, promoting reflection, and shaping people’s behavior. In a similar spirit, Jo focuses on offering personalized feedback based on users' journal entries to assist their reflection and actions towards their well-being.  

The Health Mashups project~\cite{Bentley:2013:HMP:2533682.2503823} aims to identify connections between well-being data and its context. It then describes significant observations and insights from users' data in natural language (e.g., {\em you walk more when you sleep better}). The results showed that such insights led to positive behavior change. Jo shows its users insights based on analyzing the activities the user performs and the life values they relate to.  

Importing activity data to make the application useful without requiring the user to enter all the data is a key challenge in well-being applications. Some research (e.g., UbiFit's Mobile Sensing Platform (MSP)~\cite{Consolvo:2008:ASW:1357054.1357335}) has looked into ways to automatically recognize people's physical activities in real time, and allow users to manually add missing activities. In a previous version of Jo we attempted to import journal moments from social media posts of the user.  We disabled this feature because we found that most users write differently for themselves than when they post for others. However, finding appropriate ways of importing users content from elsewhere remains an important direction to explore.

In addition to the academic research work, there has been recent interest to develop products that help users incorporate the findings of the science of happiness into their daily lives. Current applications that pursue this goal generally fall into one of the two categories: (1) applications that suggest relevant content to the users based on their answers to a predefined set of questions\cite{Killingsworth,Happify,Happier} or (2) applications in which users can log their emotions in a journaling-style environment but that content is available mostly for their own reflection\cite{Bliss,Mojo,DayOne,Daylio}. In both cases, the applications do not provide personalized feedback to the user in a way that helps them plan their future actions.

\section{An overview of Jo}
\label{sec:jo-design}
Users of Jo journal their experiences with the app. Jo then tags their moments based on a variety of criteria and presents the user several ways of reflecting on their past experiences. Jo also provides feedback to the user that helps them set and achieve certain goals. Finally, Jo helps the user create reminders that can result in additional positive experiences. 

Jo  incorporated design strategies for technologies that support behavior changes in everyday life, summarized by Consolvo et al.~\cite{Consolvo:2009:TDS:1518701.1518766}. For example, following the ``positive strategy", we used positive reinforcement in Jo to encourage users. Following the ``controllable strategy", we always let users add, edit, delete their data.

In what follows, we describe Jo's features in more detail and explain the choices we made.

\subsection{Journaling}

\begin{figure}[hbt]
\begin{center}
    \includegraphics[width=0.35\textwidth]{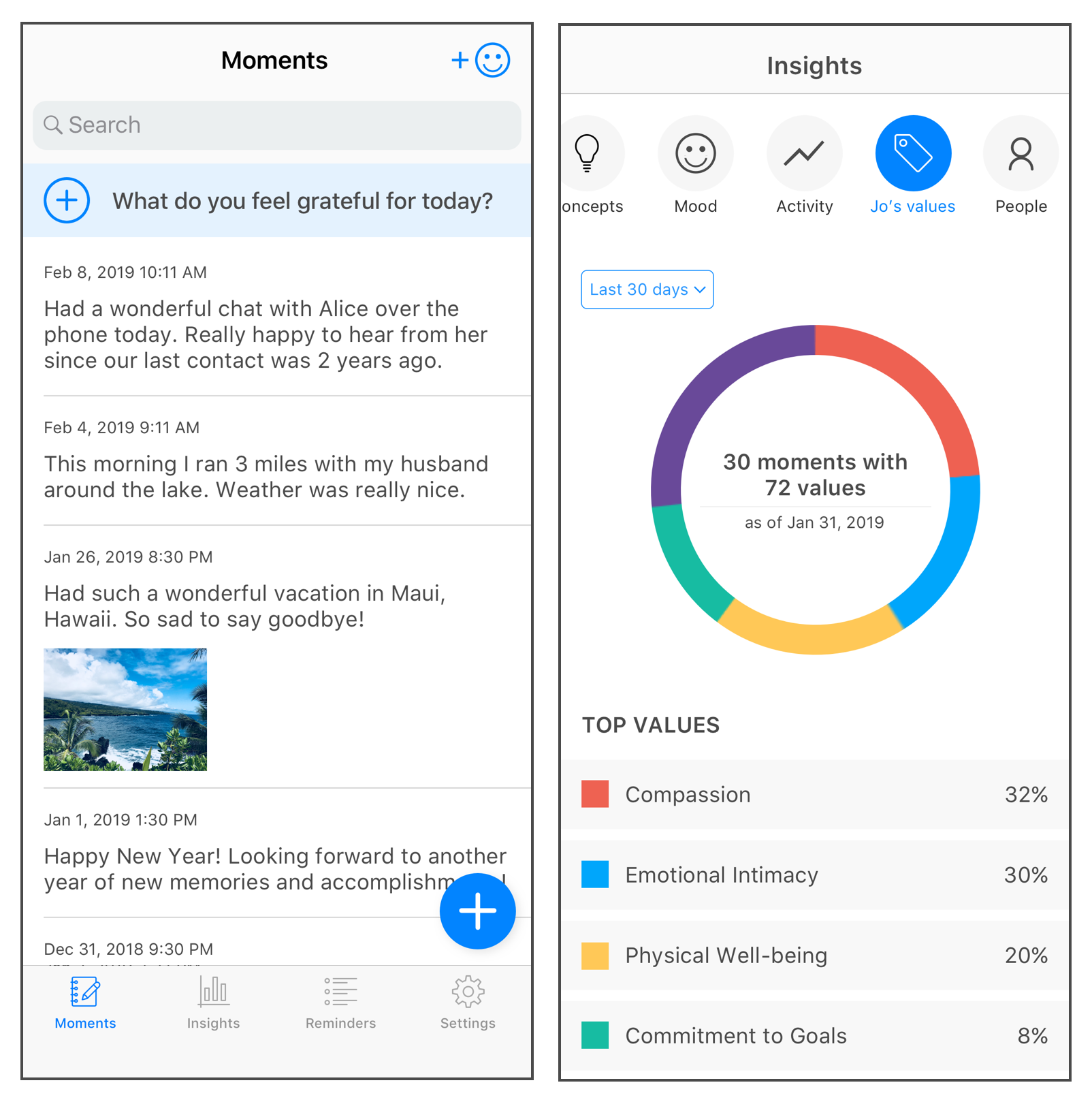}
\end{center}

\vspace{-2mm}
\caption{Reflecting: the screen on the left shows the timeline of users journal entries. The screen on the left shows a pie chart visualization of the top ranked values mentioned in users moments}
\vspace{-3mm}
\label{fig:reflect}
\end{figure}

Users log short journal entries into Jo, which we often refer to as moments (see the left hand side of Figure~\ref{fig:reflect}). We chose journaling as the primary method to capture the user's daily activities because it is a straightforward interaction. Journaling is superior to having the user answer survey questions (in the spirit of the DRM~\cite{drm}) because it captures the nuances of the user's experience and doesn't bog her down with irrelevant questions. Another interaction method we considered is a conversational interface, but  conversational technology requires future work to be applied here. 

Journaling on a mobile app provides easy access, which allows users to log experiences throughout the day. Furthermore, research has shown that  negative experiences  have more impact on people than positive ones~\cite{nass2010,Baumeister2001,Amabile2001}. Hence, journaling right after having a (possibly fleeting) positive moment also reduces bias in recall.  In our experiment, the participants reported that they like the convenience of using the smartphone compared to using the traditional pen and paper. 

Jo users log both positive and non-positive experiences. Users are encouraged to log significant moments they experience throughout the day. In the first version of Jo, we instructed users to write down only positive moments, with the goal of encouraging them to pay more attention to the positive moments in their lives.
However, we received strong user feedback that suggests that understanding what contributes to their unhappiness is as important as knowing happiness contributors. Furthermore, knowing how to navigate through negative emotions improves one's well-being. As a result, we decided that Jo should capture all experiences that are meaningful to the user. As we describe in Section~\ref{sec:nlp}, Jo identifies automatically whether a moment is a positive or negative one.

Initially, users reported that they run out ideas for moments to journal. To assist the user, Jo provides a set of prompt questions which it chooses randomly and displays at the top of the screen they use to enter a moment. Examples include ``What do you feel grateful for today?" and ``What made you laugh today?"
Our experiments found that these prompts are very useful. In particular, users showed greater levels of gratefulness based on being prompted about it. 
 
\begin{figure}[hbt]
\begin{center}
    \includegraphics[width=0.3\textwidth]{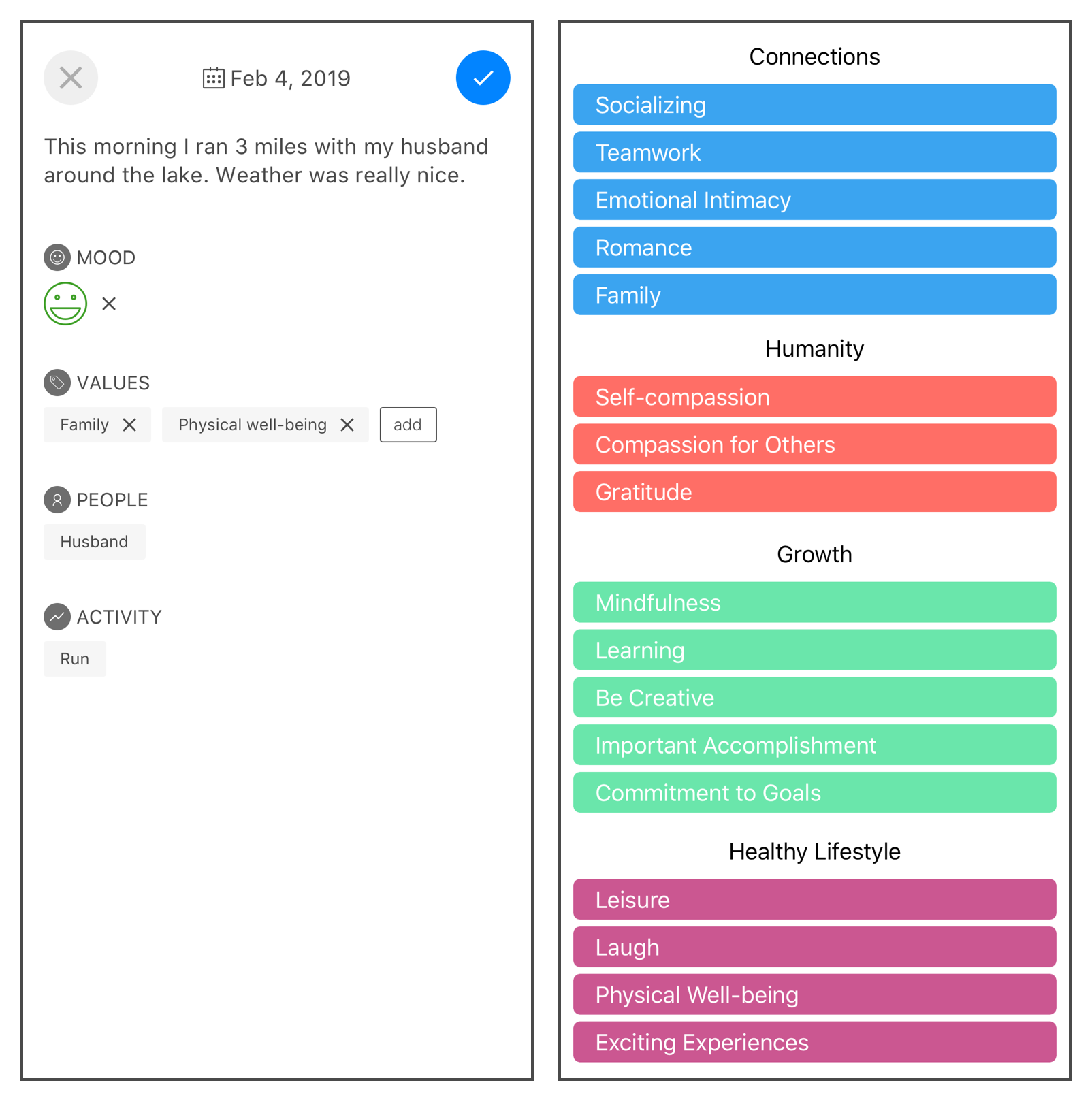}
\end{center}
\vspace{-2mm}
\caption{Tagging: the screen on the left shows an example of tags assigned to a journal entry, including type of experience, values, people and activities. The screen on the right shows a sample of list of Jo's values.}
\vspace{-3mm}
\label{fig:values}
\end{figure} 

\subsection{Tagging moments with values}

Jo attaches a set of tags representing {\em life values} to each moment the user enters (Figure~\ref{fig:values}). The concept of values comes from Positive Psychology, where they are meant to represent a set of principles an individual holds at one's core. Jo used 17 values that are commonly important to the general population, falling in into the following categories:  Connection (Socializing, Teamwork, Emotional Intimacy, Romance, Family), Humanity (Self-compassion, Compassion for others, Gratitude), Growth (Mindfulness, Learning, Be creative, Important accomplishment), and Healthy Lifestyle (Leisure, Laugh, Physical well-being, Exciting experiences). These 17 values are chosen from a set of 88 values from the Value-Based Living literature.

The tags are used in Jo's visualizations and suggestions. The user is free to add or remove tags. We describe how we map moments to tags in Section~\ref{sec:nlp}.

\subsection{Reflecting}

People who  have low self-awareness and struggle with self-reflection have trouble enhancing  their well-being and growth~\cite{Leary2018}.
Hence, one  of the key goals of applications that promote well-being is to raise the awareness of the user.  There are two areas one can cultivate to improve emotional awareness. The first is to encourage the person to pay attention to their experiences. The second area is to understand the correlations between events and emotions.

Jo offers users two ways to reflect on their experiences, shown in Figure~\ref{fig:reflect}.
First, in the Moments screen, users can view a timeline of journal entries in detail, with the newest ones on top. The photos that were attached to the moments are also shown to make the timeline more compelling. Users can also search for specific moments with keywords or values. This feature allows users to reminisce about their experiences, offering a general view of their life. 

The second place users can reflect is on the Insights screen, where they can see a dashboard summarizing their experiences and emotions. The summary is organized by key components extracted from the user's positive experiences, with pie charts that visualize top ranked values, people and activities. Users can reflect on such insights, and develop awareness of what contributes to their happiness. 

The user can also take a more active step toward working on the values they want to enhance by setting goals. Users can choose three values they would like to focus on, and set a goal for the number of moments to enter in a week that embody these values. Jo counts the number of moments that are associated with each of these three values, using a progress bars to track the goal completion. The user can use this insight to guide their actions, and decide how to spend time. 

\begin{figure}[hbt]
\begin{center}
\includegraphics[width=0.35\textwidth]{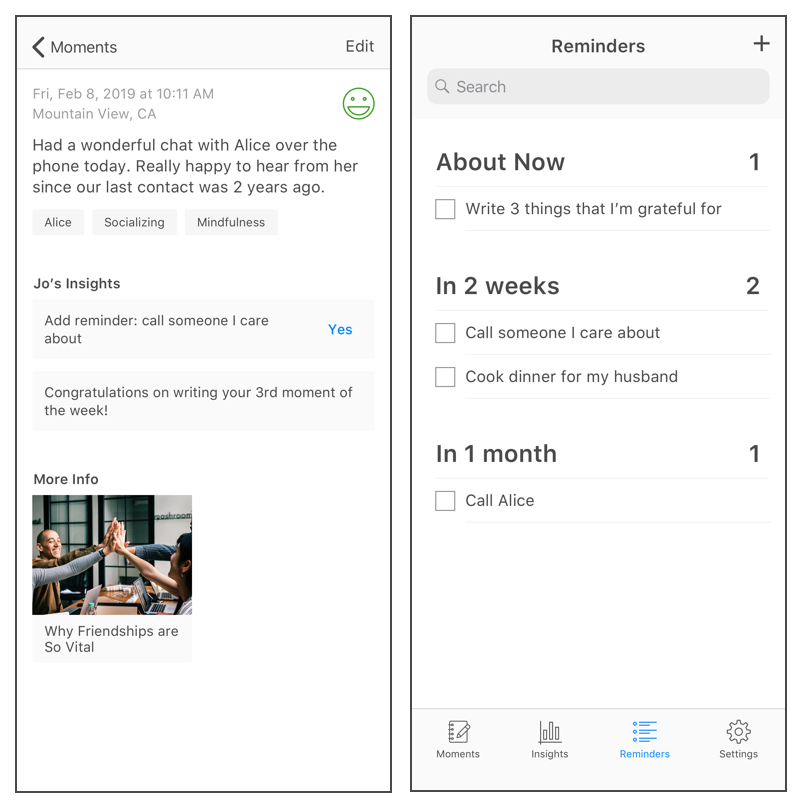}
\end{center}
\vspace{-2mm}
\caption{Responding and Reminding: the screen on the left shows Jo's responses after users enter a moment. The screen on the right shows the reminders of the user's want-to-do list.}
\vspace{-3mm}
\label{fig:quadrants}
\end{figure} 

\subsection{Feedback from Jo}

Unlike traditional journaling, which passively records users' entries, Jo provides feedback to the user's input   which assists them to better  understand their experiences. In our user interviews, we found that users showed more motivation to enter moments when they are congratulated with the progress they have made.

There are three main ways Jo offers feedback: (1)  short status reports (and congratulations) on goals achieved and activities completed, (2) pointers to additional reading about values, and (3) suggestions of similar activities that would likely make the user have a positive experience. 

For activities related to physical exercise, Jo counts the number of occurrence of these activities throughout the week. For example, if a user is a regular runner, who runs 3-5 times a week, Jo would report ``You ran three times during this week." In addition, Jo congratulates users when they achieve a significant milestone regarding their goals. For example, Jo would say, ``Congratulations that you  completed your weekly goal for Mindfulness!" These quick statistics are a casual way for users to help engage in their experiences. 

For cases where users enter a moment that is tagged with one or more values, Jo provides an article that discusses one of the tagged values in depth. We found that users are often unfamiliar with the values that Jo tags (or more commonly, with their full breadth), and these readings provide a convenient way of helping the user find other ways to perform activities that serve that value. For example, when a user enters ``I enjoyed the beautiful foliage around the park", the moment is labeled with Mindfulness. The user then sees a suggestion from Jo to read an article titled ``Five ways mindfulness improve your well-being.", and has the choice to read it immediately or later. 

For cases where users enter a moment that was a positive experience, Jo suggests a similar activity. This suggestion provides an idea for a future activity that is likely to lead to a positive experience. 

For each value, we curated a pool of three activities that embodies the value. 
When the user enters a moment, based on the value of the moment, Jo suggests one activity from the value's activity pool. For example, when a user reports ``I had a great time playing frisbee with my kids in the park", Jo tags the moment with the Family value, and suggests ``Set a time to cook a family meal", as the latter might also contribute to the Family value. Clearly, activity suggestions based only on tags are too coarse and should be more personalized. 
 
 \subsection{Reminding}
 
 Jo's reminders organize the want-to-do list that includes activities that would increase the user's well-being. As we described previously, users add activities suggested by Jo based on similarity to the positive experiences they entered. In addition, users can also directly add activities on this screen. 

Jo uses a rough timeline to organize the user's want-to-do list. As users add activities to this list, they also enter a time window in which they would like to act on the activity. Jo calculates the rough timeline based on the time users entered, and groups future activities into rough time periods, such as ``About now",`In a week", ``In two weeks", and ``Next month". We chose to arrange activities in a rough timeline because we want to de-emphasize the specific time and focus on the intention of what to do. We do not want these reminders to be another to-do list that would burden the user. By using rough time periods, users can still know what to do and how to plan ahead, but without feeling that they need to complete the task by a particular date.

This ready-to-use list has several advantages in assisting users to take action towards one's well-being. First of all, the list is pre-curated by the user, so the user is likely to find the activities on the list practical and realistic. Moreover, the users can use the list as an idea repository, instead of having to make decisions on the spot, which could be challenging. Lastly, the list allows the user to choose an activity that fits to the circumstance whenever the user finds some time, allowing the user to be spontaneous and flexible. 

\section{Natural Language techniques in Jo}
\label{sec:nlp}
    
The features of Jo described above require  natural language understanding techniques for tasks such as recognizing activities, mapping moments to values, and predicting polarity of moments. Developing these techniques is especially challenging because moments in Jo describe a broad range of domains and because there is very little training data initially. In some cases we relied on HappyDB~\cite{2018arXiv180107746A}, a corpus of 100,000 replies to the question: {\em what made you happy in the last 24 hours?} While HappyDB gives some insights into the language used in happy moments, it may not be the same as what people use in Jo moments. In what follows we briefly describe how we addressed some of the NLP challenges in Jo.

\subsection  {Extracting activities from Jo moments} 

In order to provide advice and suggestions for the time line,  Jo needs to recognize what activities are  described in a journal entry. However, given the breadth of activities, recognizing every kind of activity accurately would be a significant extension current NLP technology.   Instead, Jo focused on recognizing three classes of activities: 
  exercise, meals with people, and interactions (e.g., conversations) with someone else. We choose these three categories of activities because they cover a large fraction of the HappyDB corpus, and because the literature in positive psychology has shown that increasing these activities can result in significant increases in peoples' well-being~\cite{lyubomirsky2008happiness}.    

We trained an intent classifier to recognize moments that involve Exercise, Meals and Conversations. Once the intent was recognized, we used an  attribute extractor to find the specific values mentioned in the moment (e.g., the people, time spent exercising, etc.).   

Jo's intent classifier is based on a convolutional neural network. To create the training set for the model, we first devised a set of keywords for each of the activity classes. For example, for the Exercise class, the keywords include run, walk, yoga, biking, weights and others. The positive examples for the training set were HappyDB moments that contain these keywords (possibly after lemmatization).   Next, we expand the positive training examples by searching the dataset for sentences that include synonyms or other terms that are semantically similar to our seed set. Lastly, we trim the expanded positive examples by removing sentence that are not relevant to the class. This trimming was carried out by building a separate seed set of keywords indicative of sentences outside of the class, and removing sentences that match against lemmatized forms of those words. As an example of the interaction between the two sets of sentences,  
 consider the sentences {\em I played football for an hour} and {\em I watched football for an hour}. Both of them mention football, but only one of them describes  exercise. Thus, we added ``watch" to our negative seed set to remove such examples. The false positive examples were taken from HappyDB.  To classify a Jo moment, we compare the confidence score from all three activity classifiers, and assign the sentence to the class with the highest confidence score. Table~\ref{table:activity} shows the accuracy of our model. 

\renewcommand{\arraystretch}{1.6}
\begin{table}[htbp]
\caption{Activity extraction precision/recall/F-1}
\begin{center}
\begin{tabular}{|c|c|c|c|c|}
\hline
\textbf{\text{Class}} & \textbf{\text{Activity Details}} &\textbf{\text{Precision}} & \textbf{\text{Recall}} & \textbf{\text{F-1}} \\
\hline
\textbf{\text{Exercise}} & run, biking, yoga & 92.4 & 85.0 & 88.5 \\
\hline
\textbf{\text{Meals}} & lunch, dinner, breakfast & 97.1 & 99.0 & 98.0 \\
\hline
\textbf{\text{Conversation}} & meeting, chat & 93.1 & 94.0 & 93.5 \\
\hline
\end{tabular}
\label{table:activity}
\vspace{-0.8cm}
\end{center}
\end{table}

\subsection {Mapping moments to values} 
Mapping moments to life values is inherently subjective, because the definition of the values and the meaning of a given moment is open to multiple interpretations.
Hence, our goal in developing this algorithm was to provide a reasonable first guess at the values, but let the users modify them easily.

We considered two approaches for mapping moments to life values:  a keyword-based approach, and a machine learning model. In the keyword based approach, we worked with psychologists and picked a list of commonly used words associated with each value. For example, the keyword list for the value Achievement includes phrases such as finish,  accomplish, win, and etc. Sentences containing these keywords or lemmatized forms of the keywords are assigned to the value Achievement. We evaluated this approach on a set of 200 moments from Jo and found that it yields at least one correct tag for   72\% of the moments.  

In the second approach, we trained a logistic regression model. We obtained the training data as follow: we first created naive classifiers using keywords. Then we display the top 3 value labels produced by keyword based classifier to crowd workers, who chose the most accurate labels for the given sentence. Not surprisingly, the classification accuracy from the machine learning model has improved to 79\% on average. However, the recall dropped significantly. Specifically, when applied to the 100k moments in HappyDB, the machine-learning classifier yielded 67,000 tags versus 380,000 tags by the keyword based approach. At the moment, Jo uses the keyword based approach.

\subsection {Detecting positive vs negative moments} Jo users enter both positive and negative moments, and therefore it is important for Jo to be able to distinguish between the two. One obvious baseline candidate to consider is a sentiment-analysis algorithm, and so we tried the Google-NLP sentiment analyzer available in the Google Cloud Natural Language service~\cite{google-cloud}. However, the service provided only 28\% accuracy. As an example of its failures, the moment {\em my best friend broke his leg} was classified as positive, and {\em I went for a run today} was classified as having neutral sentiment. 
 As these examples show, Google-NLP  tends to be too conservative because the language in Jo moments can convey positive and negative experiences in  subtle ways.
 
 To address the challenge, we augmented  Google-NLP  with a model that was trained directly on positive and negative moments. We trained a BiLSTM classifier with 941 positive moments from Jo, and 156  negative moments that were taken from Jo and from SimplyConfess.com, a site that is used to express negative experiences. Jo first applies Google-NLP, and if it yields negative sentiment, it stops. Otherwise, Jo feeds the moment to the BiLSTM classifier. The combined classifier resulted in 92\% accuracy on a test set of 100 moments.\footnote{The test set includes 50 positive moments from Jo and 50 negative moments from online forums such as Reddit.com.}

\section{Evaluation}
\label{sec:evaluation}
We conducted two studies to evaluate the effectiveness of Jo. The main goal of the evaluation was to assess the design choices we made and test whether users find Jo useful. 
While we tracked several psychological metrics during the evaluation, a full study of whether Jo improves well-being would require a longer study and is beyond the scope of this paper. 


\subsection{Experimental setup}

We conducted the studies with students at local universities. The participants were students taking psychology classes (but were not necessarily psychology majors) because they would likely to have an interest in well-being and related technologies.  The first study (33 students, fall, 2018) was used mostly to get feedback on the basic design of Jo, which was incorporated into the version used in the second study. 

The second study was conducted in the winter semester of 2019. We chose to launch the study at the beginning of the semester when the pressure of academic work is relatively low and students could spend more time thinking about their well-being. We recruited a total of 39 students, 13 males and 26 females, with diverse ethnic backgrounds (13 identified as white, 8 as Hispanic or Latino, 12 as Asian, 1 as African American, and 5 as other or bi-racial). The age distribution included 30 participants in the 18-24  range, 8 in 25-35, and 1 was over 65.  A bit more than half (52\%) of the participants used self-help tools (apps, books) in the 3 months leading up to the study.

Since Jo was available only on iOS, we first surveyed which kind of smartphone they use (31 out of the 39 students had
iPhones). Then we proceeded as follows. We randomly chose a participant. If they didn't have an iPhone, we put them in the control group. If they did, we randomly assigned them between the experimental or control group. We continued until we filled the experimental group, which ultimately included 20 out of the 39 participants. Note that each group had a consistent experience internally independent of the smartphone they used.

 The participants in the experimental group were instructed  to use Jo for 4 weeks and enter 3-5 moments a day.  The control group  was instructed to use SimpleNote,  an application that offers a 
 minimal journaling experience without any of the smart features. The participants in the control group were asked to   write 3-5 of daily experiences in a checklist format (adapted from~\cite{Lyubomirsky:2011}) for the same period.
 
We collected a wide variety of measures related to well-being through an extensive survey at beginning and end of the study, including measures having to do with gratitude, depression, mindfulness and life satisfaction. We also conducted discussions in class at the end of the experiment and in depth interviews with the 4 users who used Jo the most.  

\subsection{Results and discussion}
Of the  39 participants in the study, 28 (72\%) completed  both the initial and final survey after 4 weeks (16 from the experimental group and 12 from the control group).  
 In terms of usage, 14  out of the 16 participants in the treatment group used Jo at least one time throughout the 4 weeks of the intervention (87.5\%).  Over the 4 weeks, the mean number of times participants used Jo was 20.87 (SD= 19.41, ranging between 0-68).

As noted and confirmed by our measurements, 4 weeks is too short to notice significant improvements in well-being of our participants. However, of the psychological measurements we collected, the measure of gratitude did show an upward trend. The gratitude survey GQ-6~\cite{McCullough2002} assesses  individual differences in  experiencing gratitude in daily life and has good internal reliability~\cite{McCullough2002}. For example, participants are asked to express their level of agreement with statements such as  ``I have so much in my life to be thankful for" or  ``Long amounts of time go by before I feel grateful to something or someone”.

We considered the difference in gratitude between the end and the beginning of the study (analyzed using SPSS/Anova). The result was a  trend-level difference between the treatment and control groups, with significance level  p = .09 (where anything below 0.1 is considered a trend and below 0.05 is considered significant), with a partial effect size of .11 (range: 0-1).   While this is a small effect, the trend is positive and the significance level would be more pronounced if shown to a bigger set of participants. This result was correlated with the feedback we got in the survey about Jo's  prompts. Jo randomly selected one of 6 prompts at the top of the journaling screen. The participants noted that the prompt   {\em what are you grateful for today?} was the most influential and nudged them to make a journal entry. Similarly, users who entering moments through prompts {\em Tell me about an experience you shared with friends} reported they started to pay more attention to meaningful interactions with friends. These results suggest  that nudging is an effective tool to get users to log moments and reflect about them, but we should consider how to tailor these prompts to the needs of the user and to vary them with time.

\smallskip
\noindent
{\bf Tagging moments in Jo:} Recall that Jo assigns values to each moment the user enters. 10 out of 15 users in the experimental group answered positively to the question: {\em Do you find that knowing the values related to your moments helps you?}
 Several themes were mentioned in the interviews: 
\begin{itemize}
 \item  Give meaning to undesirable tasks. {\em Before I used Jo, finishing math homework seemed like a task that I  need to complete. But Jo gives this moment a meaning that I am learning and I am doing great.} 
 \item  Raise awareness on personal values. {\em It showed me what I was good at valuing and what I don’t value enough.}; {\em It explains feelings and motivation.} 
\end{itemize}
The  users who did not find value tags useful claimed that they were obvious. We found that  value tags were most useful for less commonly discussed values, such as compassion, mindfulness and gratitude.  


\smallskip
\noindent
 {\bf Tracking goals:} To assess the effectiveness of goal setting as a mechanism to improve well-being,  the survey asked   {\em Do you find setting goals helps you take more actions towards the values you choose?}
11 out of 15 users reported positively (sometimes and frequently). Some reasons that surfaced from the survey answers  include: 
\begin{itemize}
 \item  Make the next step clear. {\em It helps break a task down into easy steps}; {\em It makes things well-organized and easy to understand my next step}
 \item  A sense of accomplishment. {\em Once I see a goal it feels good to do it and check it off a list}; {\em Goals always help motivate}.     
  \end{itemize}

The surveys and interviews also highlighted a few other observations about Jo that should be used as a guideline for future developments. 

\smallskip
\noindent
{\bf Explain insights in words:} Participants expressed a desire to have insights explained qualitatively. They want to see narratives that explain Jo's charts, questions that guide them to learn about themselves and explanations for why certain values contribute to their well-being or which values they should invest in more.

\smallskip
\noindent
{\bf Improve daily reminder:}  Jo sends a daily notification at 8pm to remind users to enter moments and to reflect on the day. Users reported that it was helpful at the beginning, but it quickly became a chore. In the discussions it came out that  a notification that reminds users earlier during the day and that varied in content would be much more useful. 



\smallskip
\noindent
{\bf Keep users engaged.} One of the key challenges with applications like Jo is to keep users engaged and to come back often. Many of the users were active during the first two weeks of using Jo, but used the app less frequently during the second half of the study. The reason that was most often quoted is the lack of instant reward. A related issue is that the activities that Jo suggests to users to improve their well-being were deemed too general. We believe that Jo needs to focus more on personalization, both to offer more specific activities and to offer insights that would keep users engaged. One obvious direction is to leverage one's location and local weather to propose activities.

\section{Conclusions}
We introduced Jo,  a mobile journaling application that provides feeback to its user to improve their well-being.  We descried Jo's design, its underlying algorithms and results from field studies. To fully realize the potential of applications like Jo requires further research. In particular, we believe that the key to further developments is to engage the user through a combination of  insightful questions (e.g., nudges to write entries), providing relevant content based on users' experiences, and discovering insights from the user's journal entries. In the longer run, by deriving user preferences from Jo, we plan to incorporate such preferences to tailor users' search experiences, and ultimately life experiences, in applications such as \cite{Yuliang2019, Evensen2019}.


\vspace{12pt}
\bibliographystyle{plain}
\bibliography{references} 

\end{document}